# SUMMARY FOR THEORY OF THE FERROELECTRIC PHASE IN ORGANIC CONDUCTORS IN RELATION TO EXPERIMENTS.

Serguei Brazovskii

LPTMS/CNRS, bat.100, Univ. Paris-Sud, 91405, Orsay, France;

Landau Institute, Moscow, Russia

e-mail: brazov@ipno.in2p3.fr

**Abstract:** New phenomenona in $(TMTTF)_2X$ compounds unify an unusual variety of concepts: ferroelectricity of good conductors, structural instability towards Mott-Hubbard state, Wigner crystallization in a dense electronic system, ordered $4k_F$ density wave, richness of physics of solitons, interplay of structural and electronic symmetries.

The ferroelectric state gives rise to three types of solitons: π- solitons (holons) are observed via the activation energy $\Delta$ in conductivity *G*; noninteger α- solitons (ferroelectric domain walls) provide the low frequency dispersion; topologically coupled combined spin-charge solitons determine *G(T)* below a subsequent structural transition of the tetramerisation. The photoconductivity gap $2\Delta$ is given by creations of soliton - antisoliton pairs.

The lower optical absorption comes from a collective electronic mode which, for the FE case, becomes the electron-phonon resonance combined with the Fano antiresonance. The ferroelectric soft mode evolves from the overdamped response at $T_0$. The reduced plasma frequency signifies the slowing down of electrons by acoustic phonons.

The effects should exist hiddenly even in the Se subfamily giving rise to the low frequency optical peak, the enhanced pseudogap and traces of phonons' optical activation.

Another, interchain, type of the charge disproportionation known in the relaxed $(TMTSF)_2ClO_4$ is still waiting for attention, possibly being hiddenly present in other superconducting cases.

*This article collects main theses for results which details and complete references can be found in* [1-4]. *The full set of viewgraphs will be available at the author's web site* ipnweb.in2p3.fr/~lptms/membres/brazov/. *Experimental plots below are the results by Monceau, Nad, et al, particularly from publications [1,3,5].*

## 1. Introduction. *History and Events.*

The discovery of first organic superconductors (Bechgard, Jerome, et al) gave rise also to a number of directions which none could preview in advance. The most known finding of the Field Induced Spin Density Waves followed almost immediately. But the second route, charge disproportionation (CD) and particularly its striking form of the ferroelectricity (FE) or sometimes Anti-FE (AFE), has been waiting for two decades. Finally it came (Monceau et al, Brown et al) to overthrow the almost reached consensus on the phase diagram (as reminded e.g. in Bourbonnais talk) and to revise some popular views on the role of electronic correlations (cf. Giamarchi talk). (See [2] for detailed history excursions, and more in talks by Brown and Seo.) As our ISCOM has demonstrated, the phenomenon is gaining space by expanding to new compounds (talks by Coulon and Ravy). Also the CD has been known already for a few years among layered organic materials (talks by Dressel, Mori and Seo) where also the FE has been observed more recently (Matsui et al).

Actually there was more than enough of warnings through decades. In the prehistoric epoch it was the structural $4K_F$ anomaly (Comes&Pouget, Kagoshima; this discovery can also be classified as a never expected one) which impact upon electronic properties has been shown via the $4K_F+2K_F$ lock-in (Jerome). The sequence of transitions in MEM-TCNQ (compare with T-ReO$_4$ below) and in modern ¼ filled compounds (Kanoda et al; talks by Coulomb and Ravy) can be viewed as the $4K_F$ (or Wigner) crystallization, as well as, in today's retrospective, cases of the CD. Already in early 80's, a very rich information has been accumulated on subtle structural transitions of anion orderings (AO) (Moret, Pouget, Ravy, et al), but these "gifts of magicians" were left beyond the attentions of the community (even by the 20 years' anniversary design of the "universal phase diagram"). As an exemption, in mid 80's, the interplay of electronic and structural properties was emphasized, also in vain till now, in the earlier construction of the universal phase diagram (S.B.& Yakovenko); its milestones were the cases of the intersite CD in T-SCN and the interchain one in S-ClO$_4$ (in contemporary language). But the most striking part of this story were the totally unattended "structureless" transitions (Coulon, Lawersonne, et al) waiting for revenge since mid 80's; and just these transitions have been found to be of the ferroelectric nature.

This rich history tells us about the necessity for reconciliation of different branches of Synthetic Metals which have been almost split since two decades. Indeed, the major success in finding the FE anomaly was due to precise low frequency methods for the dielectric permittivity $\varepsilon$: designed by Nad and Monceau for pinned CDWs, they were applied "illegitimately" also to SDWs in organic conductors, and finally to structureless transitions. On the theory side, the author's approach of the Combined Mott-Hubbard state [1-4] has been derived from a similar experience (N.Kirova and S.B.) in a model of some conducting polymers (curiously, the model implied just the form of the hypothetical design by Little (Jerome, introduction to ISCOM).

Below we shall discuss relations of theory and experiments in (TMTTF)$_2$X (T-X) compounds with some excursions to (TMTSF)$_2$X (S-X) ones.

## 2. Comments on experiments.

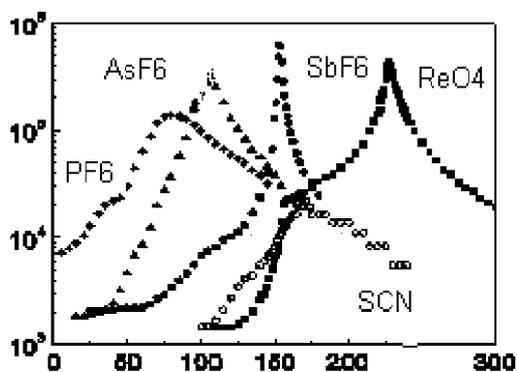

**Figure 1**. Log($\varepsilon'$(T)) at f=1MHz. FE anomaly is strikingly sharp even in the logarithmic scale. Smoothened anomaly in PF$_6$ case correlates with its weak frequency dispersion: Is it a pinning of domain walls and a hidden hysteresis? Other FE cases show the pure mono-domain "initial" FE susceptibility. The subsequent tetramerization transition in ReO$_4$ case shows up as a kink at 150K. The anti-FE case of T-SCN shows only a kink as it should be.

Remind (Javadi et al) that at GHz frequencies the X=SCN compound showed, oppositely to these low frequencies, the strongest response, with respect to FE cases; hence the purely 1D regime of the FE was recovered. Here lies also the key to the AFE/FE choice. Indeed, for highly polarizable units, the dominating Coulomb forces will always lead to an AFE structure. Oppositely, relatively weak dipole moments in compounds with centrosymmetrical ions allow for the lattice energy to make its choice in favor of the **Q**=0 FE structure.

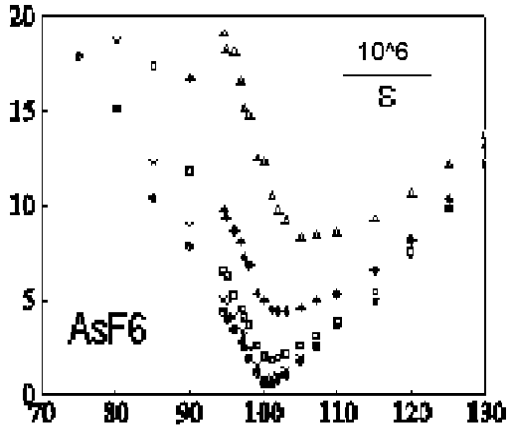

**Figure 2.** $1/\varepsilon'(T)$ at 10000, 3000, 1000, 300, 100 KHz. Apparently this is the frequency dependent depolarization of the mono domain conducting FE instead of the usual long time hysteresis. Still abundant normal carriers screen the FE polarization at the surface which eliminates the need for the domain structure. Truly low temperature studies, are necessary to find the remnant polarization. Radiational damage or other disorder will also help to pin the domain walls and fix the polarization.

**3. Solitons seen in most common cases.**

We introduce the charge phase $\varphi$ as for $2K_F$ CDW/SDW~ $\cos(\varphi+x\pi/2a)$,
then the $4K_F$ CDW~ $\cos(2\varphi+x\pi/a)$.
The two types of dimerization give rise to two sources for commensurability and to two contributions $U_s$, $U_b$ to Umklapp interactions:

$H_U = -U_s \cos 2\varphi - U_b \sin 2\varphi = -U\cos(2\varphi-2\alpha)$ ; $U=(U_b^2 + U_s^2)^{1/2}$, $\tan 2\alpha = U_s/U_b$.

$U_s \neq 0$ originates the FE ground state if the same $\alpha$ is chosen for all stacks. The state is a two-sublattice AFE if $\alpha$ signs alternate as in the $(TMTTF)_2SCN$ and more complex patterns have been found in new compounds (Ravy talk). For a given $U_s$, the ground state is still doubly degenerate between $\varphi=\alpha$ and $\varphi=\alpha+\pi$. It allows for phase $\pi$- solitons – the holons with the charge $e$ (rem. Auban-Senzier talk for new evidences).
The drawing below illustrates the origins for the potential $U\cos(2\varphi)$ by showing the sequence of zeros and extrema $\pm 1$ of the electron wave function $\Psi$ taken at $E_F$. The upper raw corresponds to correct ground states, energy minima at $\varphi=0, \pi$. They differ by signs $\pm$ which degeneracy gives rise to solitons.

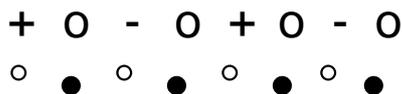

For $\varphi=0, \pi$   $\Psi=\pm 1$ at good sites ○   and $\Psi=0$ at bad sites ●   (as shown)
For $\varphi=\pm\pi/2$   $\Psi=0$ at good sites ○   and $\Psi=\pm 1$ at bad sites ●   (wrong state)

Spontaneous $U_s$ itself can change the sign between different domains of the FE displacements. Then the electronic system must also adjust its state from $\alpha$ to $-\alpha$. Hence the domain boundary $-U_s \leftrightarrow U_s$ will concentrate the non integer charge $q=2\alpha/\pi$ per chain. These $\alpha$- solitons form a gas of quasiparticles at $T>T_0$ and organize themselves in plane domain walls at T<$T_0$ (Teber et al). The essential nonsymmery in the frequency dependence of ε above and below $T_0$ (Figure 2) might be just due to this transformation.

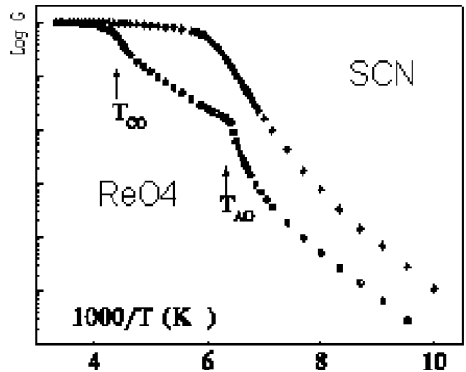

**Figure 3** shows the conductance G(T) (see more cases in [5] ). Typical plots resemble old data (e.g. Javadi et al) , but with more insights today:
1) Clear examples for conduction by charged spinless solitons: there are no gaps in spin susceptibility; also there is no difference in G(T) between FE cases and the AFE one of SCN.
2) Coexistence of both types of transitions, charge disproportionation and anion ordering; the last one increases the conduction gap and opens the spin gap.

3) Gap contribution of the spontaneous site dimerization develops very fast, and soon it dominates over the bond dimerization gap (if the last is seen at all – remind that the FE anomaly extends by >30K above $T_0$ where it can already workout a pseudogap).

Indeed, what may look as the enhanced gap (the tangent for the Ahrenius plot) near $T_0$, actually seems to be its T dependence expected as $\Delta(T)=\sqrt{\Delta_b^2 + CT_0(T_0 - T)}$, (remind, it is the IInd order transition) where the constant C happened to be large. Apparently, the differential plotting like $\Delta(T)=-d\text{Log}G/d(1/T)$ would be helpful.

So marginal effect of the build-in bond dimerization opens the route to the new compound (EDT-TTF-CONMe$_2$)$_2$X with equivalent bonds (rem. talks by Coulon and Ravy). Its ordered state has been already identified (Ravy talk) as a complex incommensurate AFE structure, showing fascinating pseudo-Ist order phase transition. This new phase is waiting to be tested for NMR splitting and ε anomaly or at least for traditional signatures of "structureless" transitions like the thermopower.

Finally, in no case there seems to be a place for weak effects of the 4-fold commensurability (Giamarchi talk) as one could have already guessed based on experience of typical ¼ filled CDWs (see more in [4]).

## 4. Effects of subsequent transitions:
### *Spin-charge* reconfinement and *combined solitons*.

Another present from the Nature is the tetramerization in (TMTTF)$_2$ReO$_4$ at $T_{AO}$<$T_0$, which leads to the spin-charge reconfinement. Similar effects are expected for the Spin-Peierls state (in T-PF$_6$) or for any $2k_F$ CDW, but the clearness of 1D regime in ReO$_4$ is uniquely suitable to keep the physics of solitons on the scenery. The ReO$_4$ plot at the Fig.3 shows, at T=$T_{AO}$, the jump in Δ (actually even in G(T), see details in [5]) which is natural for the Ist order transition. We argue that the new higher Δ at T<$T_{AO}$ comes from special topologically coupled solitons which explore both the charge and the spin sectors. Within the reduced symmetry, the invariant Hamiltonian becomes

$H_U$= -Ucos(2φ-2α) - Vcos(φ-β)cos θ

Here θ is the spin σ phase, such that θ'/π is the smooth spin density. The new, at T<$T_{AO}$, V- term is the amplitude of the mixed (β≠0 due to inversion symmetry breaking already below $T_0$>$T_{AO}$) site/bond CDW. Its formation destroys the spin liquid which existed at $T>T_{AO}$ on top of the CD. Major effects of the V-term are the following:

a) to open the spin gap $2\Delta_\sigma$ corresponding to creation of the triplet pair of new
   $\{\delta\theta=\pi, \delta\varphi=0\}$ purely spin solitons,
b) to prohibit former $\delta\varphi=\pi$ charged solitons, the holons,
   – now they are confined in pairs bound by spin strings,
c) to allow for combined spin-charge topologically bound solitons $\{\delta\varphi=\pi, \delta\theta=\pi\}$
   which leave the Hamiltonian invariant.

For the last composed particle (c), the quantum numbers are like for the normal electron: the charge e and the spin ½, but their localization is different: the spin is widely extended which is challenging e.g. for future NMR studies.

Below we suggest some schematic illustrations for the effect of the tetramerization.

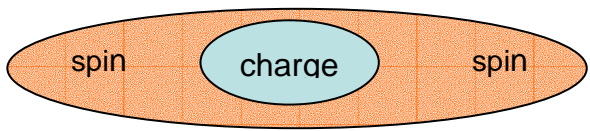

Distributions for the combined soliton:
charge e, $\delta\varphi=\pi$, sharply within $\hbar v/\Delta$
spin ½, $\delta\theta=\pi$, loosely within $\hbar v/\Delta_\sigma$

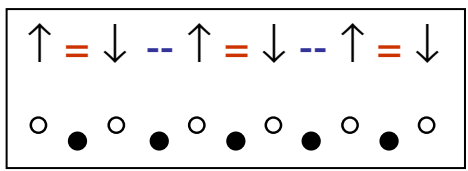

Inequivalence of bonds **=** & **--** between good sites ° endorses ordering of spin singlets. Also it prohibits translations by one ° • ° distance which were explored by the $\delta\varphi=\pi$ soliton. But its combination with the defected unpaired spin ($\delta\theta=\pi$ soliton which shifts the sequence of singlets) is still allowed as the selfmapping.

### 5. Optics: collective and mixed modes, solitons.

Remind shortly main expected optical features [4].

**I.** *In any case of the CD*, for both FE or Anti-FE orders, we expect
   (based on standard facts from theory of phase soloitons):
   Ia) Strongest absorption feature comes from the phase mode,
       analogy of the exciton as the bound kink-antikink pair;
   Ib) Two-particle gap $2\Delta$ (photoconductivity e.g.) lies higher,
       it is given by free pairs of $\pi$-solitons=kinks;
   Ic) Spectral region $\omega_t<\omega<2\Delta$ may support also quantum breathers –
       higher bound states of solitons.

**II.** *In case of the FE order* we expect additionally or instead of I:
   IIa) Fano antiresonance at the bare a phonon mode coupled to the CD;
   IIb) Combined electron-phonon resonance at $\omega_{0t}>\omega_0,\omega_t$ which substitutes for Ia;
   IIc) FE soft mode at $\omega_{fe}$
       (overdamped near $T_0$, at $T<T_0$, it increases with the FE order parameter).

Unfortunately, the optical picture of TMTTF compounds is complicated by multiple phonon lines filling just the relevant region. But this obstacle is not in vain, being viewed also as another indication for the CD. Indeed, surprisingly (kept noticed since 80's, Jacobsen et al) high intensity of molecular vibrations in TMTTF may be just due to the inversion symmetry lifting by the CD. Less pronounced phonon lines in TMTSF case may tell in favor of a fluctuational regime of the CD; it agrees with a pseudogap, rather than the true gap, in electronic optical transitions.

Hidden existence of the CD and local FE, at least in fluctuating regime, is the fate of the TMTSF compounds and the major challenge is to detect it. We argued in [1-4] that the signature may had been already seen in optical experiments (Degiorgi et al). Key effects of anion orderings (particularly the opportunity to compare relaxed and quenched phases) also are waiting for attention. Alternatively, it is tempting to try for some low $T_0$ cases of TMTTF's where the gap can fall below the mess of molecular vibrations like it happened fortunately in TMTSF's. The suppression of the CD by pressure (Nagasawa presentation, Kanoda et al) may be very helpful as well.

## 6. Conclusions and perspectives.

The ferroelectricity discovered in organic conductors, beyond its own virtues, is the high precision tool to diagnose the onset of the charge disproportionation and the development of its order. The wide range of the FE anomaly ($T_0 \pm 30K$) tells that its developing dominates the whole region below and even above these already high temperatures. Even higher are the conduction gaps, up to 2000K, forming at lower T. Remind also the TTF-TCNQ with its ever present $4K_F$ fluctuations. All that appears at the "Grand Unification" scale, which knows no differences with respect to interchain couplings, anion orderings, ferro- and antiferroelectric types, or between *Sulphur* and *Selenium* subfamilies, between old faithful weakly dimerized compounds and the new quarter-filled ones. Hence the formation of the Electronic Crystal (however we call it: *disproportionation, ordering, localization or Wigner crystallization of charges; $4K_F$ density wave, etc.*) must be the starting point to consider lower phases and the frame for their properties.
On the theory side, the richness of symmetry defined effects of the CD, FE, Anti-FE and various AOs (see [2] for earlier stages) allows for efficient qualitative assignments and interpretations.
The story of hidden surprises may not be over. Another, interchain, type of the charge disproportionation known in the relaxed $(TMTSF)_2ClO_4$ (see more in [2]) is still waiting for attention, possibly being hiddenly present in other superconducting cases. Otherwise we are forced to accept two different types of superconductivity in $(TMTSF)_2ClO_4$ and $(TMTSF)_2PF_4$. The coexistence of the CD and AO transitions in $(TMTTF)_2ReO_4$ discovered with the help of the ferroelectric anomaly warns for this possibility for other cases as well.

We conclude with the overview of main challenges: hidden existence of CD/FE in the metallic *Se* subfamily; optical identification of gaps and soft modes; physics of solitons via conductivity, optics, NMR; FE hysteresis, relaxation, domains; more exploration of AOs and other structural information.

**Acknowledgements:**
Collaboration with P. Monceau and F. Nad,
Hospitality of the YITP at the Kyoto University,
Support from the INTAS grant 2212.

REFERENCES:
[1] Monceau P., Nad F. and Brazovskii S., Phys. Rev. Lett. **86**, 4080 (2001)
                               *or* **cond-mat**/0012237.
[2] Brazovskii S., **cond-mat**/0304076                        (ISCOM 2001).
[3] Brazovskii S., Monceau P. and Nad F., **cond-mat**/0304483    (ICSM  2002).
[4] Brazovskii S., **cond-mat**/0306006                        (ECRYS 2002).
[5] Nad F.   J. Phys. France IV, **12**, 133 (2002)              (ECRYS 2002).